\documentstyle[12pt,aps]{revtex}

\include{psfig}

\input epsf

\begin{document}
\title{
\hfill\parbox[t]{2in}{\rm\small\baselineskip 14pt
{~~~~~JLAB-THY-99-21}\vfill~}
\vskip 2cm
Comment on\\ ``Valence QCD:  Connecting QCD to the Quark Model" \\~
}

\author{Nathan Isgur}
\address{Jefferson Lab, 12000 Jefferson Avenue,
Newport News, Virginia  23606}
\maketitle

\vspace{1.0 cm}

\begin{abstract}

		I criticize certain conclusions about the physics of hadrons drawn from a
``valence QCD" approximation to QCD.

\bigskip
\end{abstract}

	Lattice QCD is not just useful as a technique for calculating
strong interaction observables like the proton mass:  it can also be used
to help us {\it understand} QCD. This is the goal of the work
described in Ref. \cite{vQCD}. Its
authors present a field theory which they call ``valence QCD" (vQCD) which
they hope can be identified with the valence quark model.  The key feature
built into vQCD is a form of suppression of Z-graphs, {\it i.e.,} of quarks
propagating ``backward in time" \cite{Zgraphs}.  The authors make sound arguments for the
importance of trying to capture the essence of the quark model in a
field-theoretic framework, and present some interesting results (both
theoretical and numerical) on vQCD.  This comment is not directed at the goals of
vQCD but rather at certain conclusions about the physics of
hadrons which the authors have drawn from their work which I consider
unjustified.  Foremost among these is the claim highlighted in their abstract that baryon hyperfine
interactions are `` . . . largely attributed to the Goldstone boson
exchanges between the quarks. . . " and not to standard one-gluon-exchange (OGE) forces \cite{DGG,IK}.

%
%
\begin{center}
~
\epsfxsize=1.8in  \epsfbox{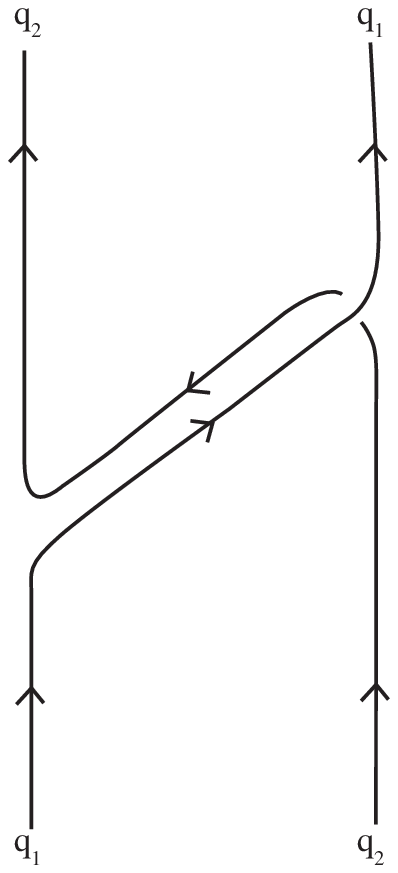}
\vspace*{0.1in}
~
\end{center}

\centerline{Fig. 1(a):  Z-graph-induced meson exchange between two quarks.}

\bigskip

%
%
\begin{center}
~
\epsfxsize=6.2in  \epsfbox{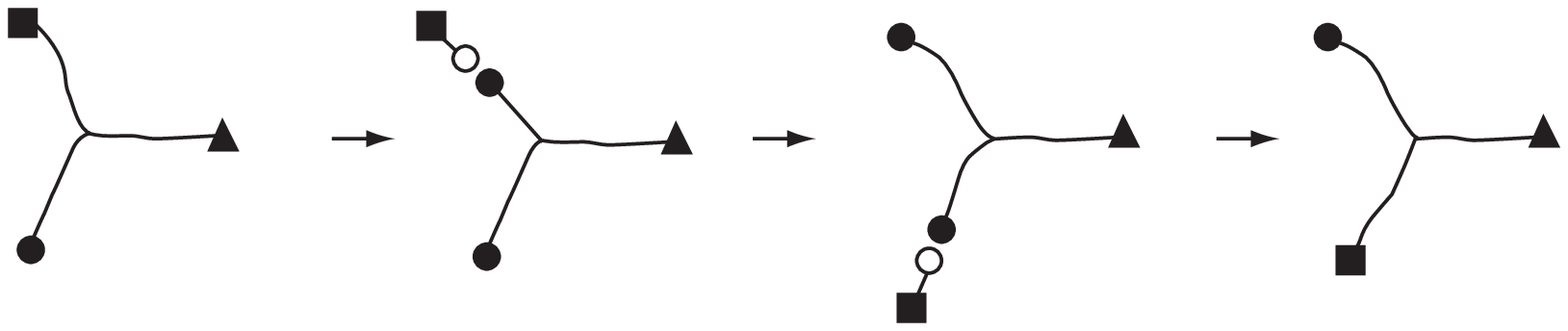}
\vspace*{0.1in}
~
\end{center}

\noindent{Fig. 1(b):  A cartoon of the space-time development of the Z-graph-induced
meson exchange in a baryon in the flux tube model.  For diagrammatic
clarity three different flavors of quarks are shown.  Note that if the
created meson rejoins the flux tube from which it originated, the produced
$q \bar q$ pair can be of any flavor; however, such a process would be a closed $q \bar q$
loop and therefore not part of the quenched approximation.  Also possible,
but not shown, are OZI-violating graphs with the creation or annihilation of
a disconnected $q \bar q$ meson; these are irrelevant to octet meson exchange in
the SU(3) limit and enter in broken SU(3) only through the $\eta-\eta'$
mixing angle.}

\bigskip

	The origin of this claim is that in vQCD the $\Delta-N$ splitting is
only about 50 MeV in contrast to the 180 MeV found on the same lattices
in standard quenched QCD (qQCD) \cite{continuum}.  The authors of Ref.
\cite{vQCD} argue that since vQCD differs from qQCD only in the suppression
of Z-graphs, Z-graph-induced meson exchange between quarks (see Figs. 1),
and in particular the exchange of the octet of pseudoscalar mesons (OPE),
must be the origin of most of the hyperfine interaction.  My objection to
this conclusion 
is that vQCD also appears to produce a very different spectrum
(and thus, one presumes, very different internal hadronic structures)
from qQCD, so that the reduced hyperfine {\it splittings} cannot necessarily
be associated with a reduction in the strength of the hyperfine {\it interaction}.
To see this we 
begin with an examination of the spectrum of vQCD, which I
have extracted from Ref. \cite{vQCD} and display in Fig. 2 in comparison
with the spectra of nature and of qQCD \cite{whya1}.    
Most masses are rough estimates based on the graphs
displayed in Ref. \cite{vQCD} and, except for the vQCD $a_1$ mass, I have
made no attempt to estimate statistical or systematic uncertainties of the
lattice ``data". However,  the problem displayed in Fig. 2 does not need such
refinements to stand out clearly:  the spectrum of vQCD is dramatically different from both
qQCD and nature!

\bigskip\bigskip\bigskip

%
%
\begin{center}
~
\epsfxsize=5.0in  \epsfbox{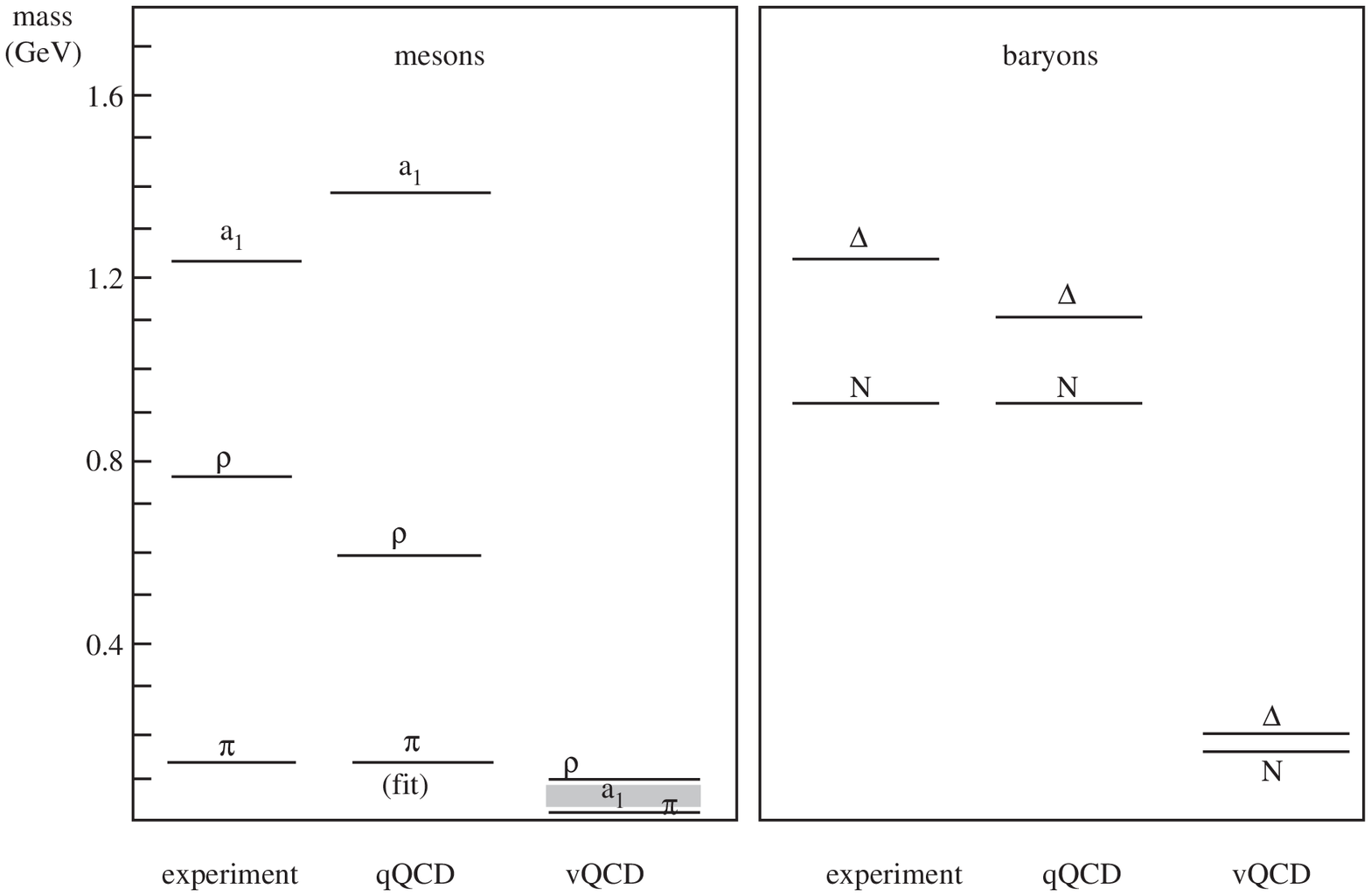}
\vspace*{0.1in}
~
\end{center}

\noindent{Fig. 2:  The meson and baryon spectra of nature, quenched QCD, and valence
QCD.  I indicate by the shaded band an estimated error for the vQCD $a_1$ mass, since
this mass is important to the arguments of the text.}

\bigskip

%
%
\begin{center}
~
\epsfxsize=6.0in  \epsfbox{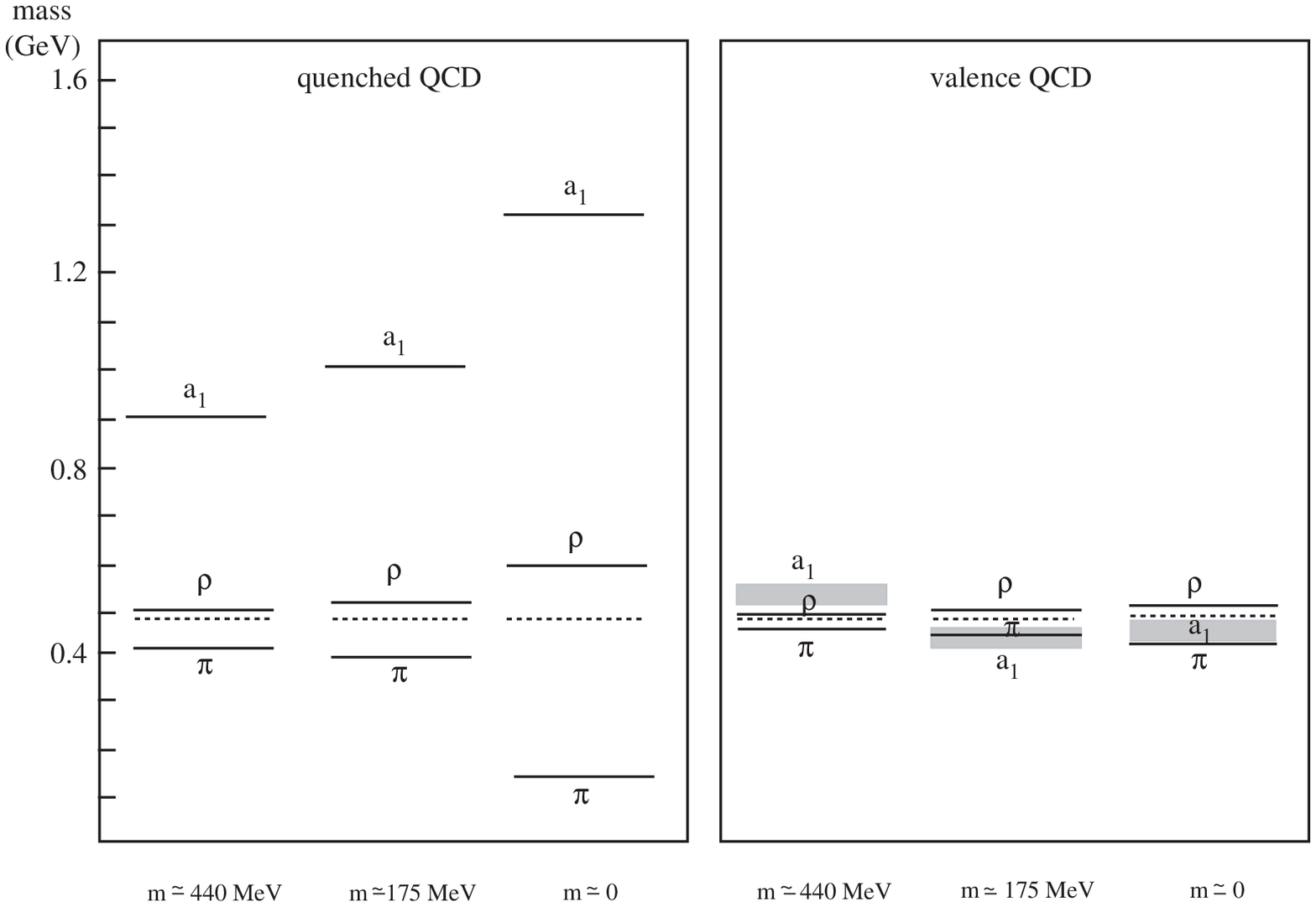}
\vspace*{0.1in}
~
\end{center}

\noindent{Fig. 3:  The meson spectra of quenched QCD and valence QCD as functions of
the quark mass $m$.  The heavy quark center of gravity 
${1 \over 4}m_{\rho}+{3 \over 4}m_{\pi}$ obtained from
qQCD at $m = 440$ MeV is shown as a dotted line and is used to define an origin
for all the other spectra, since it is the $a_1$ excitation energy relative
to this center of gravity that is of interest here.}

\bigskip

	The authors of Ref. \cite{vQCD} argue quite cogently that the physical
mechanisms which produce a constituent quark mass $m_{const} \simeq 300$ MeV in
qQCD will be suppressed in vQCD, and proceed to analyze their results
assuming that their spectra will be repaired by an overall shift of meson
and baryon masses by $\sim 2m_{const}$ and $\sim 3m_{const}$, respectively.  With this
picture in mind, they conclude that the $\Delta-N$ hyperfine interaction in
vQCD is much too small and are led to the conclusion quoted above. 
However, it is clear from Fig. 2 that such
a simple shift of the spectral zeros cannot fix the vQCD spectrum:  they
would still have essentially zero orbital excitation energy (the $a_1-\rho$
mass difference).  At one point in the paper the authors mention the
possibility of repairing the vQCD spectra by ``tuning" the quark mass, though they never
discuss this option.  I have extracted their meson spectra for heavier quark
masses and plotted them in Fig. 3 for both qQCD and vQCD.  The qQCD spectra
are very reminiscent of the experimental spectra shown in Fig. 4, but vQCD
appears to be very different even for relatively heavy quark masses.  There is
certainly no indication that quark masses $m \sim 300$ MeV, which fix the spectral
zero problems in the mesons and baryons as expected, fix the problem of the
vQCD excitation spectrum. The authors of Ref. \cite{vQCD} are silent on this matter.

%
%
\begin{center}
~
\epsfxsize=3.25in  \epsfbox{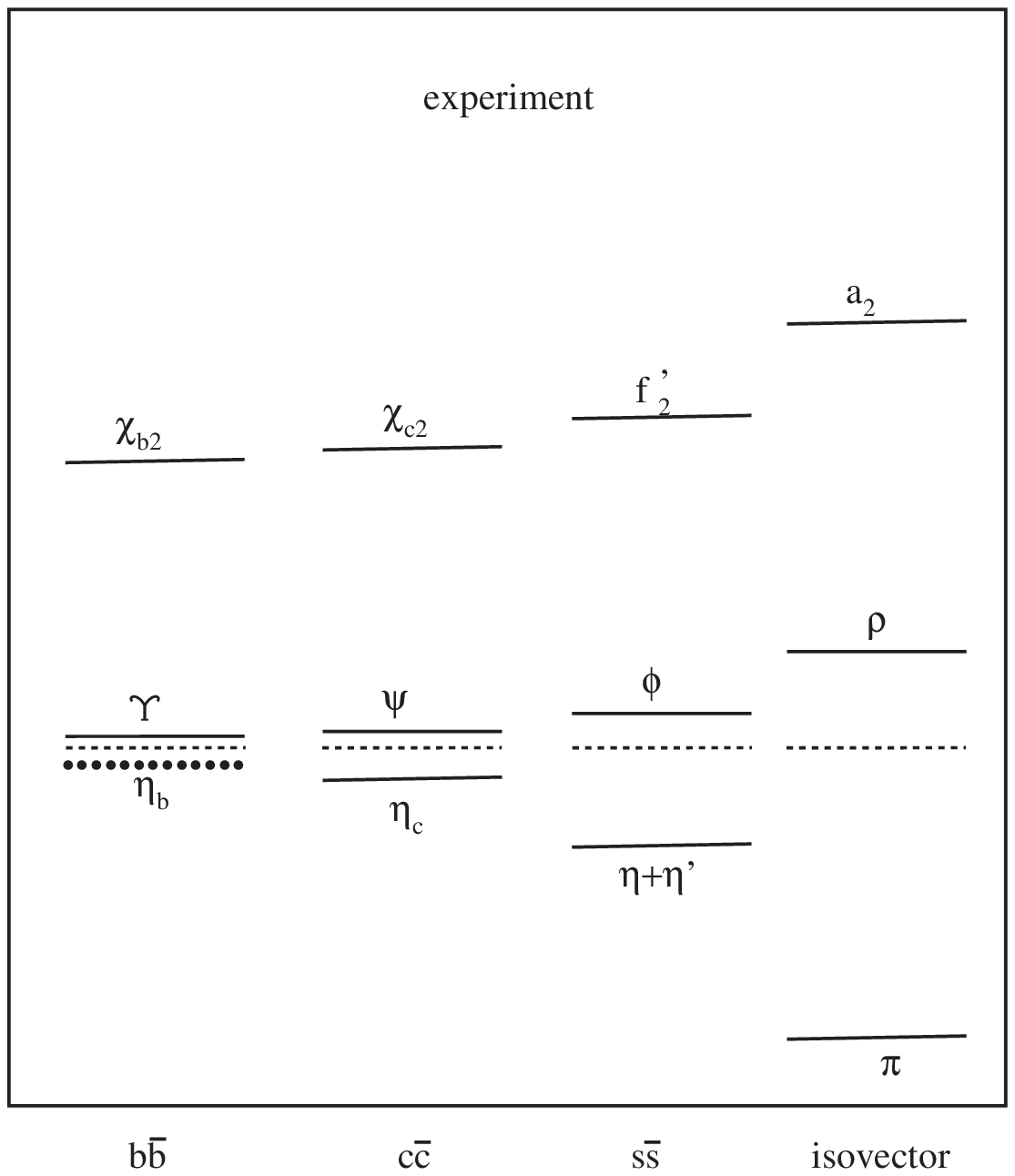}
\vspace*{0.1in}
~
\end{center}

\noindent{Fig. 4: The experimental spectra of $b \bar b$, $c \bar c$, $s \bar s$, and isovector light
quarkonia with the center of gravity of the $S$-wave mesons aligned.  
The $2^{++}$ states have been used
in lieu of the $1^{++}$ states because $\phi_1$ is not yet clearly identified. 
The pseudoscalar $s \bar s$ state (``$\eta+\eta'$") has been located by
unmixing a $2 \times 2$ matrix assumed to consist of primordial 
$s \bar s$ and ${1 \over \sqrt 2}(u \bar u+d \bar d)$ 
states.  The $\eta_b$ is not yet discovered, but the theoretical
prediction is shown as a dotted spectral line. The spectra are 
shown to scale, which may conveniently be calibrated from 
the $\chi_{c2}-\psi$ splitting of 459 MeV.}

\bigskip

	There is a complication in completing my criticism of Ref.
\cite{vQCD}.  The preceding discussion is focused on the meson spectrum
because it is only for mesons that the orbital excitation spectrum is given
in Ref. \cite{vQCD}.  Fig. 2 certainly gives the impression that vQCD has
similar problems in both the meson and baryon sectors:  in both the masses
drop precipitously and the hyperfine interactions are much weakened.
However, it is a logical possibility that the zeroth-order baryon and meson
spectra have very different dynamical origins and that the baryon spectrum
would be satisfactory after an overall shift (or tuning of $m$).  In fact,
even before showing us spectra, the authors of Ref. \cite{vQCD} show us
lattice data on vQCD which indicates that the nucleons are of normal size,
a result which suggests that the 
baryons are normal \cite{example}.  It would therefore
certainly be interesting to know where the negative parity baryons are in
vQCD.  Let me note, however, that if the zeroth-order baryon and meson
spectra have different physical origins, it would contradict much that we
think we understand about QCD and would render such apparent empirical
confirmation of that understanding as the equality of the slopes of meson
and baryon Regge trajectories into misleading accidents.   {\it If this
were  true, it would represent a far more profound conclusion than the
ones the authors of Ref. \cite{vQCD} draw:  it
would destroy rather than just modify the quark model}.  If vQCD
eventually leads us to this result, it would thus make the other physics conclusions of
Ref. \cite{vQCD} irrelevant.  The simplest possibility is surely that 
the baryon excitation spectrum is as different from qQCD and nature as was the
meson spectrum.

	In summary, the spectrum of
vQCD does not appear to represent a good approximation to nature, to qQCD,
or to the quark model.  
This does not mean that we can't learn many things from it, but it
does mean that one must be very cautious in drawing 
conclusions from vQCD.
In particular,
hyperfine splittings are normally especially sensitive functions of the
internal structure of the states being perturbed ({\it e.g.}, in
the nonrelativistic quark model they are proportional to the square of
the spatial wave function at zero separation). There is certainly no
reason to believe that a ground state system belonging
to such a poorly described spectrum will have reasonable short distance matrix elements. 

	         I would
like to point out that independent of the basic objections I have raised to
drawing physics conclusions from vQCD at this stage, the association of the
missing hyperfine strength of vQCD with Goldstone boson exchange (OPE) between
quarks suffers from some serious problems. The first of these is apparent
from Fig. 2: while qQCD describes both the $\rho-\pi$ 
and $\Delta-N$ splitting, they are both
poorly described in vQCD. It would be natural
and economical to identify a common origin for these problems.  However,
the Z-graph-induced meson exchange of Fig. 1(a) can only act between two
quarks and {\it not} between a quark and an antiquark.  Of course it is
logically possible that there are different mechanisms in operation in the
two systems but, as we will see, this is very difficult to arrange.

    Figure 4 showed the evolution of  quarkonium
spectroscopy as a function of the quark masses.  In heavy quarkonia ($b \bar b$ and
$c \bar c$) we know that hyperfine interactions are generated by one-gluon-exchange (OGE)
perturbations of wave functions which are solutions of the Coulomb-plus-linear
potential problem.  I find it difficult to look at this diagram and not see
a smooth evolution of the wavefunction (characterized by the slow evolution
of the orbital excitation energy) convoluted with the predicted $1/m_Q^2$
strength of the OGE hyperfine interaction.

	This same conclusion can be reached by approaching the light
quarkonia from another angle.  Figure 5(a) shows the evolution of ground state heavy-light
meson hyperfine interactions from the heavy quark limit to the same
isovector quarkonia shown in Fig. 4.  In this case we know that in the heavy quark limit
\cite{IW} the hyperfine interaction is given by the matrix element of $\vec \sigma_Q \cdot \vec B/2m_Q$,
consistent with the OGE mechanism and
the striking $1/m_Q$ behaviour of the data on
ground state splittings as $m_Q$ is decreased from $m_b$ to $m_c$ to $m_s$ to $m_d$.

	    Since the OPE mechanism cannot contribute here, the
OGE mechanism is therefore certainly the natural candidate for generating all meson hyperfine
interactions.  The problem that  arises for the OPE hypothesis is that it is then nearly
impossible to avoid the conclusion that OGE is also dominant in baryon
hyperfine  interactions:  the OGE $q \bar q$
and $qq$ hyperfine interactions are related by a simple factor of $1/2$, and
given the similarities of meson and baryon structure (for example, their charge
radii, orbital excitation energies, and magnetic moments are all similar),
it is inevitable that the matrix elements of OGE in baryons and mesons are
similar.  This connection is explicitly realized
in quark models \cite{universal}.

%
%
\begin{center}
~
\epsfxsize=5.5in  \epsfbox{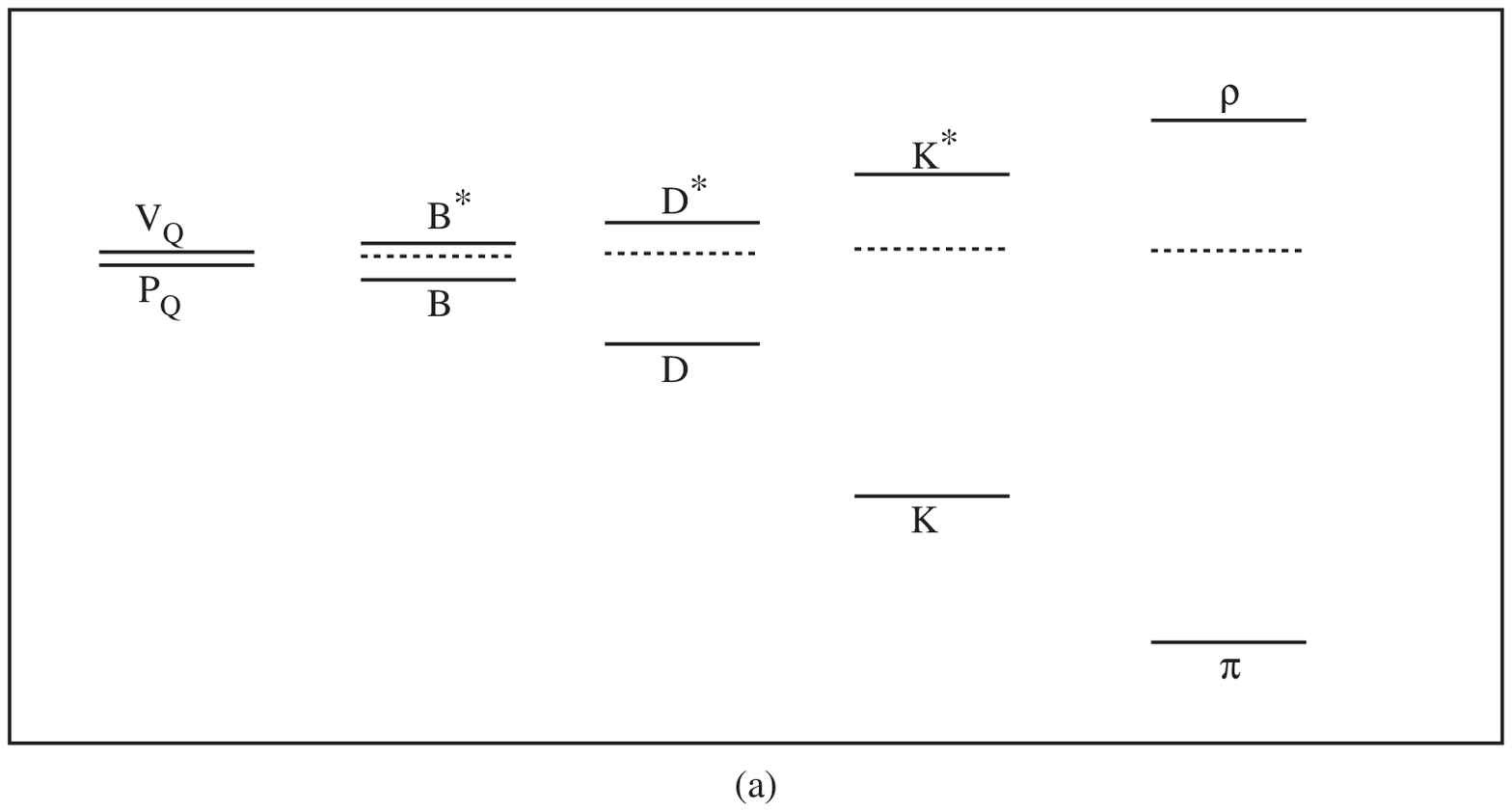}
\vspace*{0.1in}
~
\end{center}

%
%
\begin{center}
~
\epsfxsize=5.5in  \epsfbox{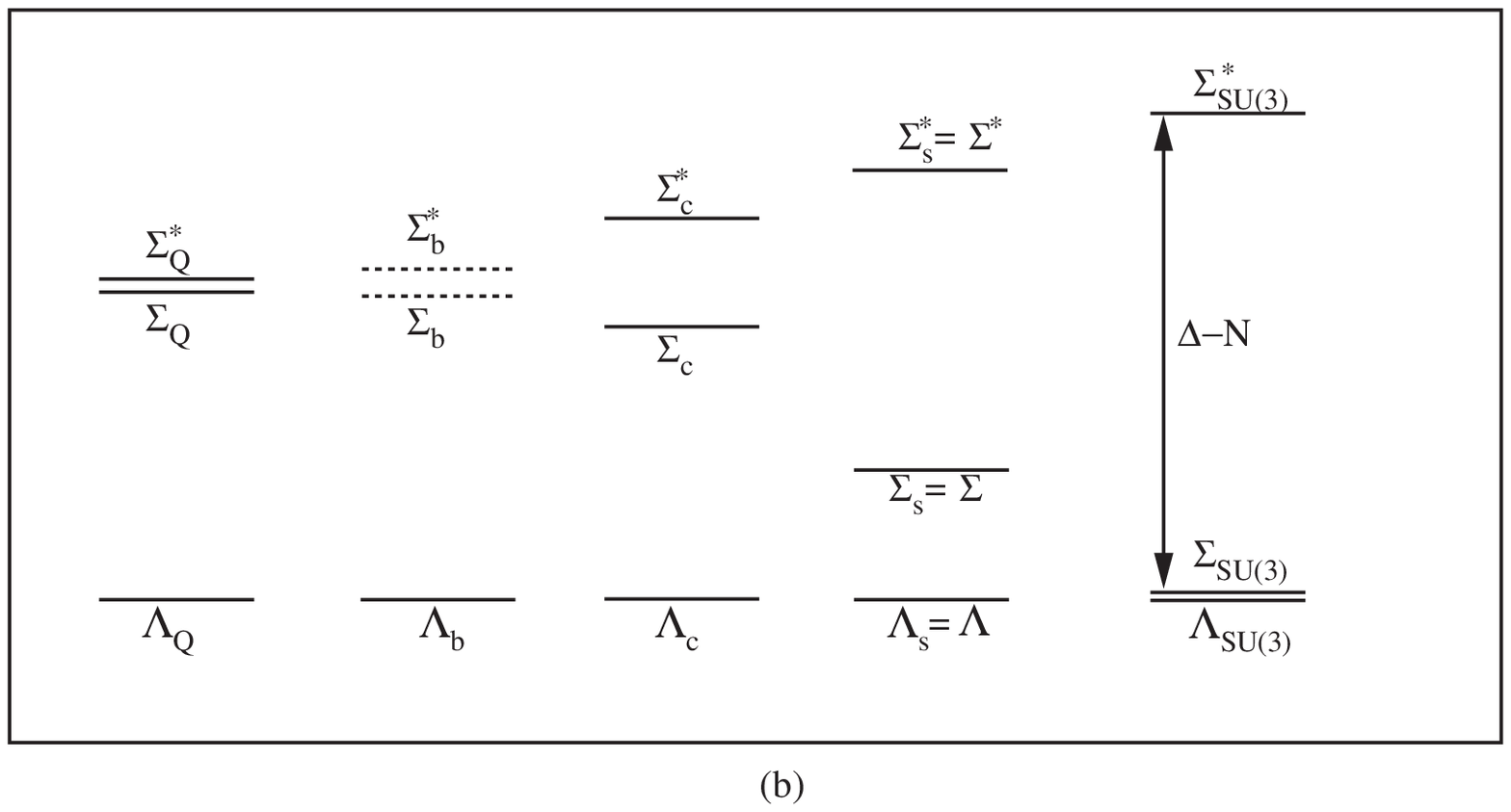}
\vspace*{0.1in}
~
\end{center}

\noindent{Fig. 5: Ground state meson (a) and baryon (b) hyperfine splittings in heavy-light
systems as a function of the mass $m_Q$ of the heavy quark. The spectra on the far left
are the $m_Q \rightarrow \infty$ limits of heavy quark symmetry. The  
$\Sigma^*_Q-\Lambda_Q$ splitting 
and the 
positions of 
$\Sigma^*_b$ and $\Sigma_b$ 
are estimates from the quark model; all other masses are from experiment.
The spectra are shown to scale; the meson scale may conveniently be 
calibrated with the $D^*-D$ splitting of 141 MeV and the baryon scale 
with the $\Sigma_c-\Lambda_c$ splitting of 169 MeV.}

\bigskip

	 There is
another very serious problem with the OPE mechanism which surfaces in
mesons.  I have explained that there are no Z-graph-induced meson exchanges
in mesons.  However,  Fig. 6 shows how the same meson
exchanges which are assumed to exist in baryons will
drive OZI-violating mixings in isoscalar channels {\it via annihilation graphs}.  
I have argued above that the structure of mesons and baryons is 
so similar that it is impossible to avoid
their having similar OGE matrix elements.  The same is true
for OPE matrix elements:  it is impossible to maintain that OPE is
strong enough to produce the $\Delta-N$ splitting in baryons without
predicting a matrix element of comparable strength associated with Fig. 6 in mesons.
Such matrix elements will violate the OZI rule \cite{OZI}.  
To see this, consider the mixing between the pure
$\omega$-like state ${1 \over \sqrt 2}(u \bar u+d \bar d)$ 
and the pure $\phi$-like state $s \bar s$.  This mixing will be
driven by kaon exchange and from the preceding very general arguments we
must expect that the amplitude $A_{OZI}$ for this OZI-violating process will
have a strength of the same order as the 200 MeV  $\Sigma^*-\Sigma$ splitting
(which is also driven purely by kaon exchange).  Empirically $A_{OZI}$ is
very tiny - - - of the order of 10 MeV - - - in all  known
meson nonets (except the pseudoscalars); amplitudes an order of magnitude larger would
lead to dramatic violations of the OZI rule.

\bigskip\bigskip

%
%
\begin{center}
~
\epsfxsize=1.0in  \epsfbox{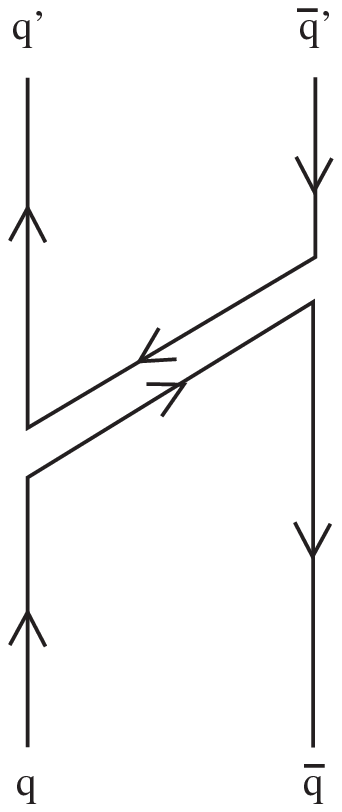}
\vspace*{0.1in}
~
\end{center}

\centerline{Fig. 6:  OZI-violating mixing in isoscalar mesons {\it via} the exchange of a $q \bar q'$
meson.}

\bigskip

	   The mesons thus produce some
disastrous conclusions for the Goldstone boson exchange hypothesis.  The first is the very
unaesthetic conclusion that two totally distinct mechanisms are in operation
producing meson and baryon hyperfine interactions:  OGE in mesons and OPE
in baryons.  The second is the virtual impossibility of having strong OGE
matrix elements in mesons without also producing strong OGE matrix elements
in baryons, in conflict with the basic hypothesis of that model. The third
is that the OPE mechanism produces unacceptably large OZI violation
in meson nonets.

	   As shown in Figure 5(b), experiment actually provides 
strong evidence  in support of the dominance of OGE and {\it not} OPE in the baryons themselves!
Fig. 5(b) is the baryon analog of Fig. 5(a) where once again one knows rigorously
that the hyperfine interactions are controlled by the matrix elements of
$\vec \sigma_Q \cdot \vec B/2m_Q$ for heavy quarks.  
It is clear from this Figure that in the heavy quark
limit the OPE mechanism is not dominant:  exchange of the heavy
pseudoscalar meson $P_Q$ would produce a hyperfine interaction that scales
with heavy quark mass like $1/m_Q^2$, while for heavy-light baryons the
splittings are behaving like $1/m_Q$ as demanded by heavy quark
theory (with which the OGE mechanism is automatically consistent). It is
difficult to look at this diagram and not see a smooth evolution of this $1/m_Q$
behaviour from $m_c$ to $m_s$ to $m_d$, where by SU(3) symmetry
$\Sigma^*_{SU(3)} - \Lambda_{SU(3)}=\Delta - N$, the splitting under
discussion here.  One might try to escape this conclusion by arguing that
between $m_c$ and $m_s$ the OGE-driven $1/m_Q$ mechanism turns off and the $1/m_Q^2$ OPE
mechanism turns on.  From the baryon spectra alone, one cannot rule out this
baroque possibility.  However, in the heavy-light mesons of Fig. 5(a) there is
no alternative to the OGE mechanism, and since the $Q \bar q$ interaction continues
to grow like $1/m_Q$ as $m_Q$ gets lighter,  so must the $Qq$ interaction.  I
see no escape from the conclusion that OGE is dominant in {\it all} ground
state hyperfine interactions.

    The preceding discussion of the generic problems of the OPE mechanism was an extended
digression designed to dampen any remaining enthusiasm for one of the highlighted conclusions
of Ref. \cite{vQCD}. It is not logically connected to and should not
distract the reader from the main argument presented in this Comment. Valence 
QCD is a potentially interesting
field-theoretic approximation to QCD from which we could in principle learn
a great deal about the physics driving hadron structure and dynamics.  My
criticisms of the 
conclusion extracted from vQCD in Ref.
\cite{vQCD} about the physics of hyperfine interactions
are based on the fact that vQCD has a very different spectrum from
qQCD and nature. It is thus unclear whether the reduced $\Delta-N$ splitting of vQCD is
due to a diminished hyperfine interaction or a change in the 
short distance structure of the hadrons.  Until it is
better understood, vQCD must only be used with great
care in drawing conclusions about the physics of hadrons.

\bigskip\bigskip

{\centerline {\bf ACKNOWLEDGEMENTS}}

\medskip

    This work was supported by DOE contract DE-AC05-84ER40150 under which the 
Southeastern Universities Research Association (SURA) operates 
the Thomas Jefferson National Accelerator Facility.

\bigskip\bigskip

{

\end{document}
\begin{references}
\bibitem{Zgraphs} It should be noted that the time-ordered graphs of vQCD are not the same as those of
the normal Feynman propagator. Among other differences related to the projector ${1\over 2}(1+\gamma^0)$,
the vQCD dispersion relation is $E=p^2/m$ versus the relativistic 
$E=(p^2+m^2)^{1\over 2}$ or the nonrelativistic $E=p^2/2m$.
\bibitem{whya1}Ref. \cite{vQCD} also gives results for the $a_0$.  The scalar
sector is so controversial and so poorly understood experimentally that I
have focused on the $a_1$, but no conclusion I have drawn depends on this
selection.
\bibitem{example} A poor spectrum does not always lead
to a poor description of the ground state. For example, the ground state of the Coulomb-plus-linear
potential problem can in certain circumstances be dominated by the Coulomb
region so that its gross properties are independent of the string tension,
leaving only  short distance effects like the hyperfine interaction sensitive to the slope of
the linear potential.
\end{references}
